# Breaking Guardrails, Facing Walls: Insights on Adversarial AI for Defenders & Researchers

Naz Bozdemir, Giacomo Bertollo, Jonah Burgess

*Abstract – ai_gon3_rogu3* was a 10-day AI red-teaming CTF co-run by HackerOne and Hack The Box in September 2025. Of the 504 registrants, 217 were active and attempted 11 jeopardy-style challenges spanning output manipulation and data-exfiltration scenarios. We analyze solve rates, engagement patterns, and tactic archetypes to understand where participants succeed and where defenses hold. Simple guardrail bypasses were nearly universal, whereas multi-step defenses created pronounced bottlenecks: the introductory task reached ~98% completion among active players, but only ~34% solved the final scenario. Output manipulation challenges saw higher success (≈82.5%) than data-extraction tasks (≈74.6%), suggesting "social-engineering the model" remains easier than extracting protected content at scale. Our findings offer a field-based complement to red teaming benchmarks, clarifying which attack families dominate in competitive settings and where resilient defenses still frustrate skilled adversaries.

## Introduction

AI red teaming brings security thinking to LLM applications by probing failure modes such as prompt injection, output manipulation, and sensitive-data exfiltration. While automated and curated benchmarks (e.g., JailbreakBench [1], HarmBench [2]) are increasingly used to test models and defenses, comparatively fewer studies analyze community-scale behavior in the wild. We study *ai_gon3_rogu3 [3]*, a 10-day competition with 504 registrants and 217 active players, to quantify solve dynamics, tactic stratification, and choke points across 11 challenges. We find sharp skill stratification, higher success for output manipulation than for data extraction, and strong effects of format-obfuscation tactics, with multi-step defenses remaining robust, among other insights.

1. **Sharp skill stratification**, nearly all active players solved the easiest challenge, but only a third managed the final scenario.
2. **Output manipulation proved easier than data extraction**, highlighting players' intuitive grasp of prompt-based misclassification.
3. **Common adversarial tactics remain highly effective**, especially encoding tricks and format obfuscation.
4. **A distinct skills gap exists**, separating a core of elite players from a broad long tail of less experienced participants.

This study gives researchers and defenders a data-driven view of how adversaries actually probe and break AI systems. It shows which attack strategies work and how quickly they spread in a competitive setting. For the CTF community, the results reveal that simple exploits were rapidly solved by nearly everyone, while more complex defenses created choke points. Only a small group of highly skilled players managed to push through those barriers, underscoring where real difficulty lies in securing AI systems.

## Background and Related Work

*Human and automated red teaming.* Early work formalized LLM-vs-LLM [4] red teaming and showed that auxiliary models can uncover safety failures at scale. More recent efforts introduced open robustness benchmarks such as JailbreakBench [1] and HarmBench [2] that standardize behaviors, threat models, and scoring. Mining in-the-wild jailbreaks (e.g., WildTeaming) [5] complements these curated sets, and many-shot jailbreaking [6] highlights long-context vulnerabilities.

*Prompt injection & agent threats.* Indirect prompt injection [7] in tool-integrated agents expands the attack surface beyond simple chats; recent benchmarks (e.g., InjecAgent) systematize such scenarios. We connect our observations to risk taxonomies, OWASP Top-10 for LLM Applications [8], and MITRE ATLAS [9], to bridge CTF behaviors with defender frameworks.

*CTFs and AI security skill evaluation.* Emerging attempts involve LLMs solving CTFs [12–14], LLMs competing against humans [11], and training LLM agents in CTF-like environments [22,23], but less is known about human solver dynamics in AI red-teaming CTFs. Our results fill this gap, providing baseline participation and solving-rate distributions to inform the design of layered defenses and training pathways.

*External testing complements CTFs.* AI-focused bug bounty programs and external testing are already surfacing jailbreaks [11, 16] and unsafe behaviors in deployed systems. Recent work argues that structured, third-party adversarial evaluations [15] should be part of model governance, and that external incentives accelerate disclosure and remediation, while classifier-based defenses offer a complementary mitigation layer [15, 16]. We view AI red teaming, CTFs, and bug bounty programs as complementary modes of scrutiny across the AI system lifecycle. Bug bounties and coordinated disclosure programs surface jailbreaks and unsafe behaviors in deployed systems, while CTFs reveal skill distributions, emergent tactics, and the diffusion of exploit patterns within structured environments. AI red teaming acts as a structured, adversarial stress test across models and configurations, probing guardrails, emergent behaviors, and exploits [17, 20], bridging the gap between bench tests and live system vulnerabilities. By combining all three, defenders can maximize coverage (breadth), depth (scenario variety), and realism (deployment fidelity) in AI security evaluation.

## Methodology

We analyzed detailed logs from the CTF platform, including every solved challenge, interaction attempt, and challenge instance launch. Our dataset:

- **Participants**: 504 registered, with 217 active players (defined as solving ≥1 challenge).
- **Metrics**: Challenge solves, completion rates across registered vs. active participants, and interaction counts (as a proxy for effort).

The dataset includes every challenge's own (flag capture) record, as well as logs of challenge instance usage. Each record contains a timestamp, user and team identifiers, and the challenge involved. For each challenge, we computed the number of solves (teams that captured the flag) and the completion rate among players, both as a percentage of all registered players and of active players. This is reported as "percentage ownership" in CTF terms, i.e. the fraction of teams that "owned" (solved) the challenge. Such metrics help normalize difficulty independent of the drop-outs. We also measured engagement through the count of challenge instances launched as a proxy for attempts, and the distribution of solves per player. Finally, we categorized challenges by type (e.g. "data extraction" vs. "output manipulation" tasks) based on their scenario and the primary skill required, to compare performance across challenge categories.

As an example, *Re-Cars AI* (the introductory challenge) had 213 solves, equating to 42.3% of all registered teams and 98.2% of active teams. In contrast, the final challenge *Performance Crossroads* saw 74 solves – only 14.7% of all teams (34.1% of active teams). We tabulated such percentages for every challenge. Across all users, we logged 2,116 environment launches and 1,772 successful flag submissions, implying ~16% of sessions did not yield a solution (344/2,116 ≈ 0.163), or about one in six. By dividing the total launches by the number of active players (2,116/217), we find an average of ~9.7 instances launched per active player, indicating that most participants attempted nearly all challenges and/or retried some challenges multiple times.

Similarly, the median number of challenges solved per active player was 10 (out of 11), reflecting that many who solved one challenge went on to solve many more (the competition had a high retention among engaged players, as shown later). The next sections will illustrate the findings.

## 2.1 Ethics

All analyses were conducted in accordance with the event's official terms of participation and applicable data protection principles. The dataset used in this study contained no personally identifiable information (PII); all participant identifiers were anonymized prior to analysis. Results are presented exclusively in aggregate form to prevent any inference about individual participants. The study's objective was limited to understanding player performance trends and aggregate behavioral patterns within the context of AI red teaming, ensuring adherence to ethical research standards and respect for participant privacy.

## 2.2 Challenge-authoring caveats (LLM-specific)

LLM CTF tasks are inherently stochastic and path-diverse [18]. Because model behavior, tool integrations, and guardrails can be non-deterministic, an "intended" solution may not always trigger consistently, and preventing clever unintended solutions is difficult. In our event, some teams solved the final challenge without using the stored "review" feature; an acceptable outcome because we graded *outcomes* (safety or data-access property violations), not a single canonical path. This motivates future work on challenge design patterns that are robust to model variability.

# Results

## 3.1 Participation and Engagement

Out of 504 registrants, 217 were active (≈ 43%). This left a majority (≈ 56%) who never solved a single challenge (Figure 1); a common CTF phenomenon. Active participants, however, engaged deeply: many achieved their first solve quickly, as shown in **Figure 2**, and over half went on to solve at least 10 challenges. Notably, one-third completed all 11 tasks, marking a high-performing elite (Figure 8).

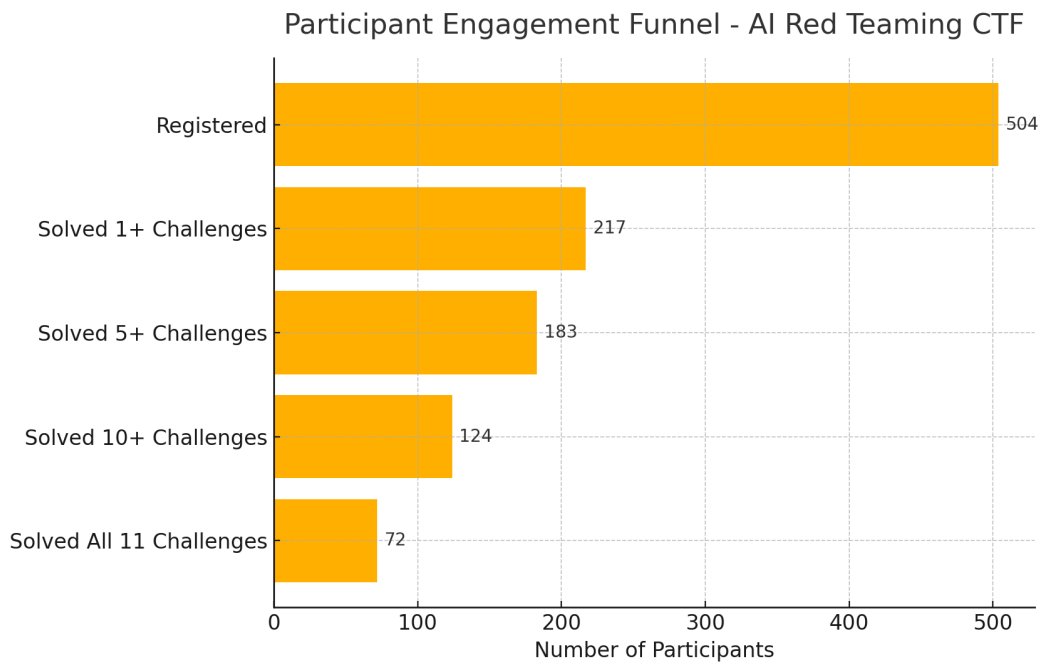

**Figure 1. Participant Engagement Overview** – Distribution of registered vs. active participants. This funnel chart illustrates participant progression throughout the event. Out of 504 registrants, 217 achieved at least one solve, and 72 completed all 11 challenges. The steep early drop-off indicates a moderate entry barrier in AI red teaming tasks, yet strong retention among active solvers.

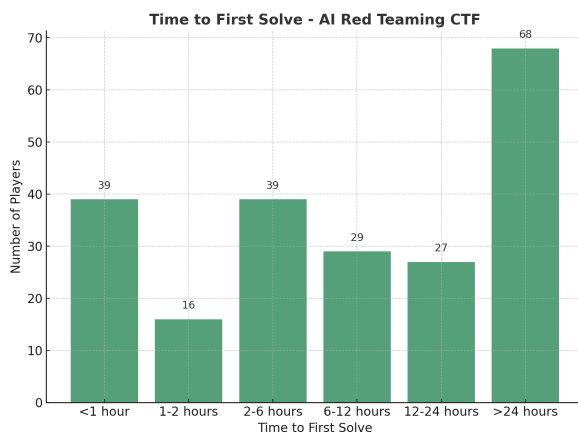

**Figure 2. Time to First Solve per Active Player –** Distribution of how quickly participants achieved their initial success. This figure shows the distribution of time taken by participants to achieve their first challenge solution. The distribution shows a large >24h group (n=68) alongside a substantial early-solve cohort (<1h: n=39; 2–6h: n=39). Rather than strictly bimodal, the pattern reflects a heavy right tail. This suggests a strong divide between experienced AI security practitioners and newcomers adapting to red teaming methodologies.

Unless otherwise stated, time to first solve (Figure 3) is measured from a team's first challenge-instance launch to its first recorded solve (removing timezone/availability effects). Using this baseline, the median time-to-first-solve was ~0.32 hours (IQR 0.16–1.21 h). For context, measuring from CTF start yields a median of 13.6 h (IQR 2.11–36.68 h) with a heavy >24 h tail—reflecting participation timing rather than problem-solving latency.

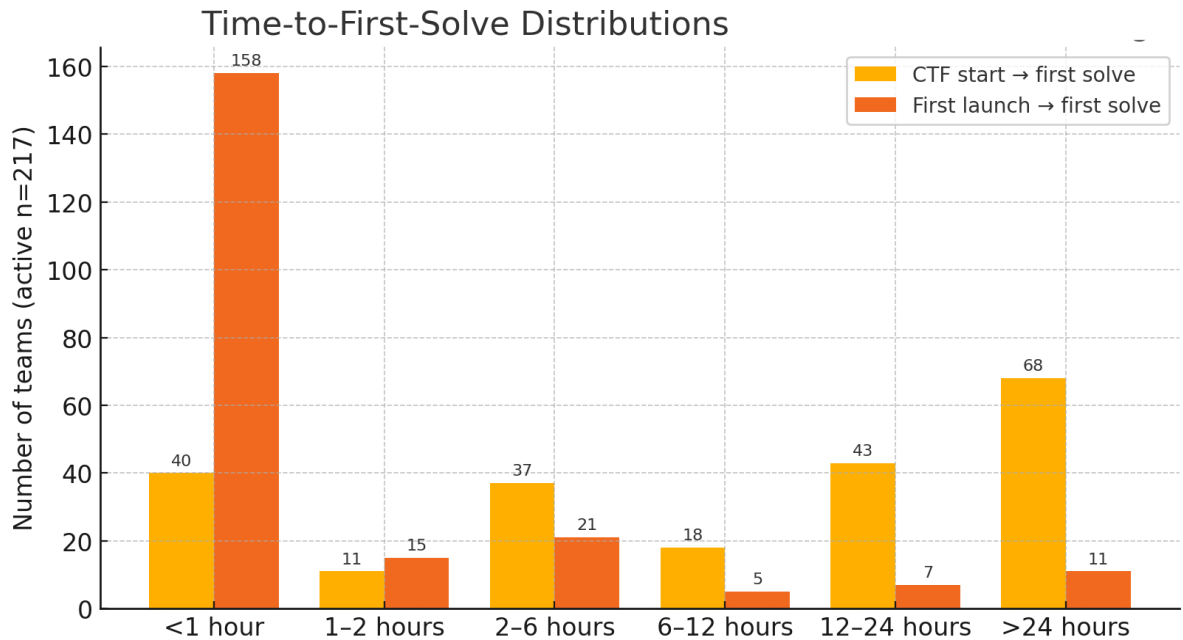

**Figure 3. Time-to-First-Solve Distributions** – Grouped bars compare counts of active teams (n=217) across time bins measured from CTF start vs. from first instance launch. Bins <1 h, 1–2 h, 2–6 h, 6–12 h, 12–24 h, >24 h have counts:

- CTF start → first solve: 40, 11, 37, 18, 43, 68
- First launch → first solve: 158, 15, 21, 5, 7, 11

The CTF-start baseline reflects availability/timezone effects, whereas the first-launch baseline isolates on-task latency.

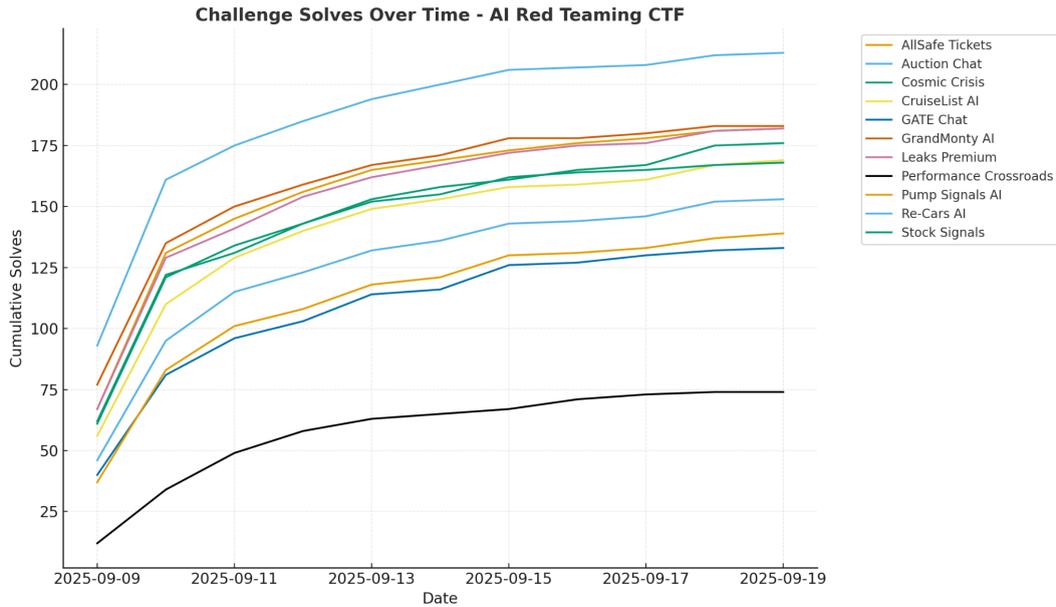

**Figure 4. Challenge Solves Over Time by Challenge –** Cumulative solves for each challenge across the event window. Early momentum is highest in the first 48 hours, followed by a gradual plateau. The top-performing challenges (*Re-Cars AI* and *Leaks Premium*) reached saturation quickly, indicating they served as effective onboarding tasks for participants.

## 3.2 Challenge Completion Rates

Figure 5 illustrates the number of teams that solved each challenge, in descending order of solves. The easiest challenge, *Re-Cars AI*, was solved by 213 teams, whereas the hardest regular challenge (prior to the final), *GATE Chat*, saw 133 solutions. The final boss, *Performance Crossroads*, was solved by 74 teams. When considering all registered teams, the drop-off is starker: only 14.7% of all teams achieved the final flag, and 26.4% solved *GATE Chat*, compared to 42.3% who solved the first challenge.

**Table 1. Challenge Completion Statistics**

| Challenge Category | Challenge Difficulty | Challenge Name | Challenge Total Flags | Teams Completing Challenge | % of ownership (across all CTF teams) | % of ownership (across active teams) |
|---|---|---|---|---|---|---|
| AI | Easy | Re-Cars Ai | 1 | 213 | 42.7% | 98.6% |
| | | Leaks Premium | 1 | 182 | 36.5% | 84.3% |
| | | Grandmonty Ai | 1 | 182 | 36.5% | 84.3% |
| | | Allsafe Tickets | 1 | 182 | 36.5% | 84.3% |
| | | Cosmic Crisis | 1 | 175 | 35.1% | 81.0% |
| | | Cruiselist Ai | 1 | 169 | 33.9% | 78.2% |
| | | Stock Signals | 1 | 168 | 33.7% | 77.8% |
| | | Auction Chat | 1 | 153 | 30.7% | 70.8% |
| | | Pump Signals Ai | 1 | 139 | 27.9% | 64.4% |
| | | Gate Chat | 1 | 133 | 26.7% | 61.6% |
| | Medium | Performance Crossroads | 1 | 74 | 14.8% | 34.3% |

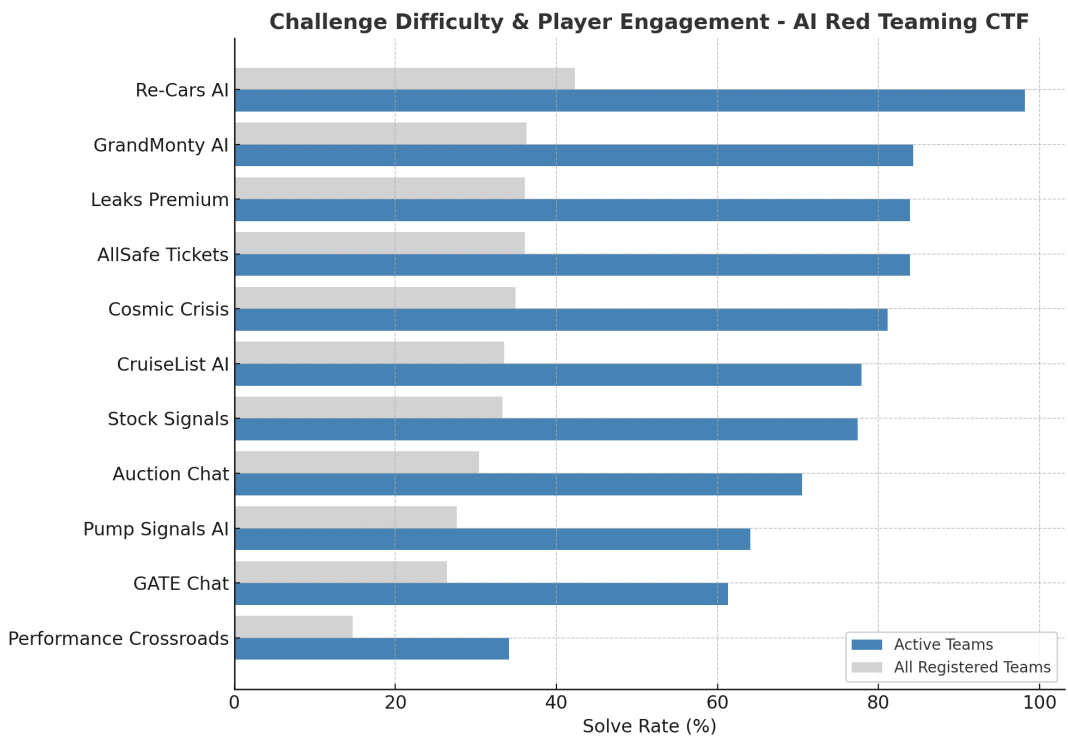

**Figure 5. Solve Rates by Challenge Difficulty** – Decline from 98.6% on the easiest challenge to 34.1% on the final boss. This summary table ranks challenges by completion rate and ownership percentage among teams. *Re-Cars AI* had the highest completion (42.7% of all teams), while *Performance Crossroads* proved the most difficult (14.8%). These results highlight the balance between accessibility and technical depth across the CTF design.

Among active teams (n = 217), 170 (78.3%) launched the final challenge (Figure 6); 74 (34.1%) solved it; and 96 (44.2%) launched but did not solve. 47 (21.7%) of active teams never attempted the final at all. Progress tiering explains this: every team with ≥10 solves (124/124) attempted the final and 73 (58.9%) solved it, whereas among teams with <10 solves (n = 93), only 46 (49.5%) attempted the final and 1 (1.1%) solved it.

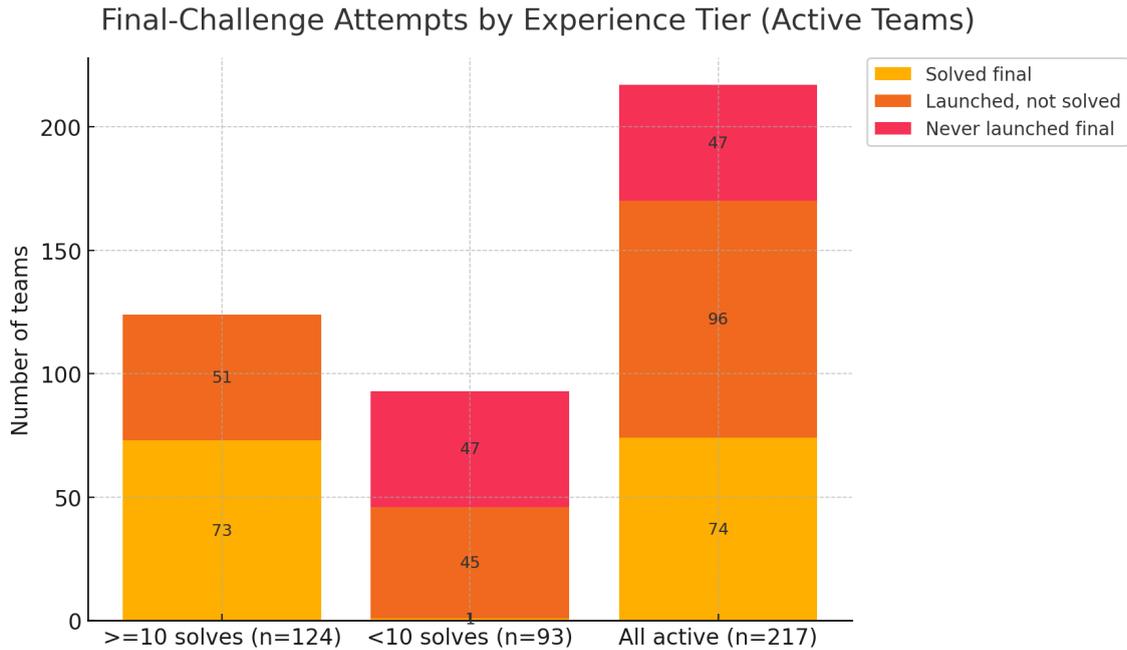

**Figure 6. Final-Challenge Attempts by Experience Tier (Active Teams)** – Stacked bars show, for teams with ≥10 solves (n=124), <10 solves (n=93), and all active teams (n=217): Solved final (73, 1, 74), Launched but not solved (51, 45, 96), and Never launched final (0, 47, 47). These counts separate *engagement* (launch) from *success* (solve) on the final challenge.

### 3.3 Category-wise Performance

Challenges fell into two broad categories:

- **Output Manipulation (2 challenges)** – average success: 82.5% of active players.
- **Data Extraction (9 challenges)** – average success: 74.6%.

**Figure 7. Category-wise Performance** – Comparison of Output Manipulation vs. Data Extraction success rates. This figure compares solve rates across two major challenge categories — *Output Manipulation* and *Data Extraction*. Participants performed better in manipulation-oriented tasks (82.5%) than in extraction-based ones (74.6%), suggesting greater familiarity with controlling model outputs than exploiting systemic data vulnerabilities.

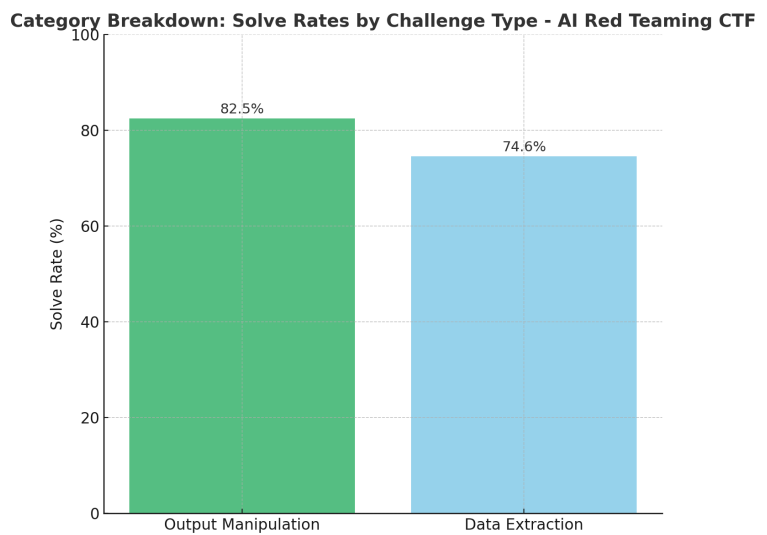

These two broad categories map onto several OWASP LLM Top 10 risks. Output Manipulation challenges align with LLM01: Prompt Injection and LLM09: Overreliance, since they demonstrate how crafted inputs can cause unsafe misclassifications that organizations may trust blindly. Data Extraction challenges map to LLM01: Prompt Injection, LLM02: Data Leakage, and LLM06: Sensitive Information Disclosure; in the case of complex scenarios like Performance Crossroads, they also touch on LLM05: Excessive Agency.

## Challenge Themes

To better contextualize the adversarial behaviors observed in the CTF, we mapped each challenge theme against the [OWASP Top 10 for LLM Applications](#) [8] and [MITRE's ATLAS framework](#) [9](Table 2).

**Table 2. Mapping CTF Challenge Themes to OWASP LLM Top 10 and MITRE ATLAS**

| CTF Theme | Challenges | Description | OWASP LLM Top 10 Category | MITRE ATLAS Mapping |
|---|---|---|---|---|
| **Basic Data Access Bypass** | Re-Cars, CruiseList | Prompt injection to override chatbot restrictions and access hidden data | LLM01: Prompt Injection | Evasion / Jailbreak – manipulating input to bypass model restrictions |
| **Gatekeeper Exploits** | GrandMonty, Auction Chat | Tricking an AI into revealing authentication or secret codes | LLM01: Prompt Injection + LLM04: Insecure Output Handling | Role Play / Jailbreak – tricking the model into acting outside its role |
| **Privilege Escalation** | Stock Signals, Leaks Premium | Extracting paywalled or premium data via crafted prompts | LLM06: Sensitive Information Disclosure | Model Extraction / Confidentiality Violation – eliciting data beyond authorization |
| **Output Manipulation** | AllSafe Tickets, Cosmic Crisis | Forcing AI misclassification (e.g., escalating ticket priority) | LLM04: Insecure Output Handling; LLM09: Overreliance | Output Manipulation / Misclassification – causing the model to produce incorrect or unsafe outputs |

| Guardrail Evasion | GATE Chat, Pump Signals AI | Defeating censorship/filters to reveal prohibited or hidden content | LLM01: Prompt Injection + LLM02: Data Leakage | Jailbreak / Censorship Evasion – defeating safety filters to access blocked knowledge |
|---|---|---|---|---|
| **Complex Multi-Step Exploit** | Performance Crossroads | Overriding role-based access controls to extract private performance reviews | LLM05: Excessive Agency + LLM06: Sensitive Information Disclosure | Multi-Stage Attack / Identity Manipulation – chaining prompts to override role-based access and exfiltrate data |

Each of these challenges highlighted different facets of AI security. From the summaries above, we see recurring patterns: many tasks required participants to bypass content filters or access controls (no direct code exploitation, since the "vulnerabilities" were in the AI's training or prompt rules). The high success rates on simpler scenarios show that techniques like formatting a query differently or using out-of-the-box phrasing are widely known and effective. Harder scenarios layered more complex guardrails (e.g., multi-turn requirements, hidden triggers), where fewer participants succeeded.

## Discussion: Key Insights

### Insight 1: A Sharp Difficulty Cliff

From nearly universal solves on the opening task to just one-third on the final boss (Figures 4 and 5), the data shows how quickly difficulty stratifies the field. This demonstrates that while baseline exploitation skills are widespread, mastering complex multi-step attacks is rare.

### Insight 2: Social Engineering the Machine

Players succeeded more often when tricking the AI into misjudging context than when extracting secrets (Figure 7). This reinforces the notion that AI security must address manipulation just as much as exfiltration risks. For educators and tool developers, this emphasizes focusing on multi-step and chained exploit training to elevate more people to that elite level.

### Insight 3: Format Obfuscation Works – Too Well

Tactics like encoding requests in JSON or base64 reliably bypassed filters. This exposes a systemic weakness: too many AI guardrails rely on pattern recognition instead of deeper semantic defenses. The community may need more practice and techniques specifically for breaking AI *security and privacy guardrails*, as opposed to just tricking AI's judgment. It also implies that current AI guardrails can be moderately effective against a large fraction of adversaries, but not all [1, 2, 16].

**Insight 4: A Bimodal Skill Distribution**

Roughly a third of players solved everything, while many solved very little (Figure 8). This gap highlights both the promise of a strong talent pool and the need for more educational on-ramps to close the divide. However, the long tail of partial completions highlights varied expertise levels in AI exploitation and prompt manipulation.

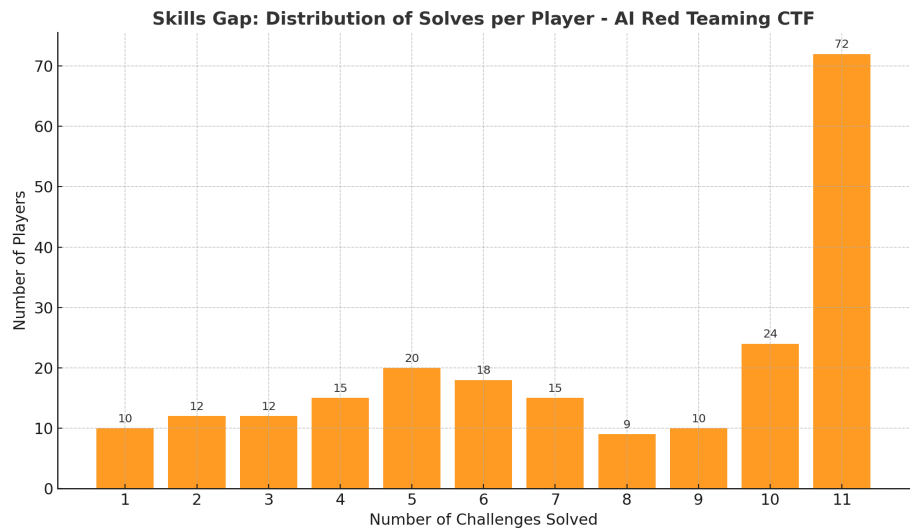

**Figure 8. Skill Distribution Across Participants** – Bimodal performance split between elite solvers and the long tail. This chart visualizes the number of challenges solved per player. The distribution is left-skewed: many participants solved all 11 challenges (n=72), with a long tail of partial completions.

# Conclusion

The *ai_gon3_rogu3* CTF revealed a community highly capable at breaking AI guardrails using known techniques but less prepared for multi-step, context-rich defenses. For defenders, the results mean that simple safeguards are trivial to bypass, but layered, adaptive protections still frustrate even skilled players, mimicking real-world attackers. For researchers, the event provides concrete data on attacker tendencies, informing the design of safer AI systems. And for the CTF community, it affirms that adversarial AI security is not just a niche interest – it's a growing frontier where skills honed today will be critical tomorrow.

---

# References

[[1] P. Chao *et al.*, "JailbreakBench: An open robustness benchmark for jailbreaking large language models," *arXiv preprint* arXiv:2404.01318, 2024.

[2] M. Mazeika *et al.*, "HarmBench: A standardized evaluation framework for automated red teaming and robust refusal," *arXiv preprint* arXiv:2402.04249, 2024.11